\documentclass[secnumarabic,aps,prl,reprint,groupedaddress]{revtex4-1}
\usepackage{verbatim}
\usepackage[english]{babel}
\usepackage{graphicx}
\usepackage{color}
\usepackage{color}

\usepackage{amsmath}
\usepackage[normalem]{ulem}
\newcommand\hl{\bgroup\markoverwith
	{\textcolor{yellow}{\rule[-.5ex]{2pt}{2.5ex}}}\ULon}

\begin{document}
	
	\title{Ideal memristor based on viscous magnetization dynamics driven by spin torque}
	
	\author{Guanxiong Chen}
	\author{Sergei Ivanov}
	\author{Sergei Urazhdin}
	\affiliation{Department of Physics, Emory University, Atlanta, GA, USA.}
	
%%\date{\today}	
	
\begin{abstract}
We show that ideal memristors - devices whose resistance is proportional to the charge that flows through them - can be realized using spin torque-driven viscous magnetization dynamics. The latter can be accomplished in the spin liquid state of thin-film heterostructures with frustrated exchange, where memristive response is tunable by proximity to the glass transition, while current-induced Joule heating facilitates nonvolatile operation and second-order memristive functionality beneficial for neuromorphic applications. Ideal memristive behaviors can be achieved in other systems characterized by viscous dynamics of physical, electronic, or magnetic degrees of freedom. 
\end{abstract}
	
\maketitle

Artificial neural networks, modeled after the functionality of brain, have recently emerged as a promising alternative to the traditional von Neumann computer architecture~\cite{hassoun1995fundamentals,neuroscience1,neuroscience2}. Among the advantages of neural networks are built-in adaptability, robust fault-tolerant operation, and the ability to process large amounts of data in real time. At the core of these capabilities is a network of neurons extensively connected by synapses. Synaptic plasticity - the dependence of the connection strength on the history of the transmitted signals - is central to the neural network's functionality~\cite{abbott2004synaptic,markram1995dendritic,markram1997regulation}.

An efficient hardware implementation of synaptic plasticity is provided by memristors - two-terminal circuit elements whose resistance is ideally proportional to the integral control parameter such as the total charge that passes through them~\cite{adamatzky2013memristor,chua1971memristor,chua2011resistance,wang2017memristors}. However, this ``ideal" memristive behavior has not been achieved yet. Instead, many types of ``generalized memristors" - devices whose properties depend in some way on their electronic history - have been explored for neuromorphic applications~\cite{wang2015overview}. 

One of the most extensively studied memristors is based on the electromigration of oxygen in metal oxides~\cite{liu2012real,strukov2008missing}. However, physical motion of atoms limits device endurance, while abrupt formation and destruction of oxygen-depleted conductive filaments results in behaviors more akin to a switch. Another notable approach utilizes magnetic domain wall (DW) motion through a ferromagnetic (F) nanowire incorporated in a magnetic tunnel junction (MTJ)~\cite{parkin2008magnetic}. The DW is driven by the spin-transfer torque (STT), and its position is read-out via the tunneling magnetoresistance (TMR). Magnetism-based operation enables high endurance, but thermal fluctuations and defects compromise controllable STT-driven DW motion at nanoscaley~\cite{meena2014overview}.

\begin{figure}
		\includegraphics[width=\columnwidth]{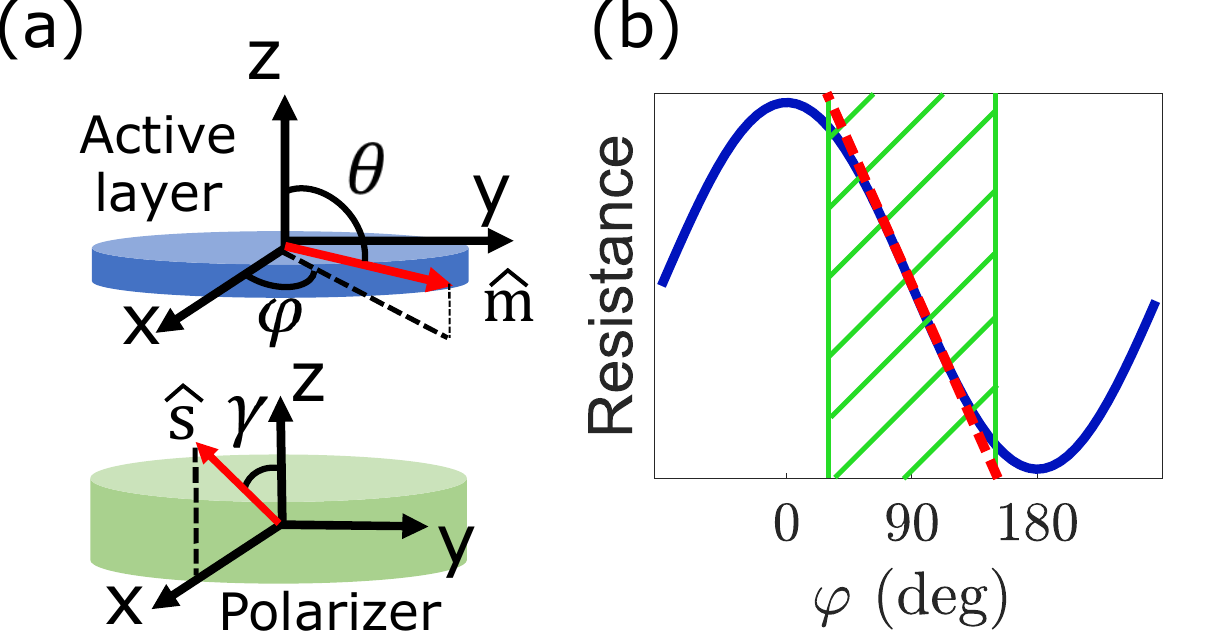}
		\caption{(a) Schematic of the proposed STT-driven memristor and the relevant parameters. (b) Typical sinusoidal dependence of resistance $R$ on the in-plane magnetization orientation. The expected angular range of nearly ideal memristive behaviors is hatched. The dashed line is a guide for the eye.}\label{fig:schematic}
\end{figure}

Here, we show that nearly ideal magnetoelectronic memristors can be implemented using nanoscale single-domain F characterized by strongly damped (viscous) dynamical characteristics, which can be achieved by sandwiching it with an ultrathin low-anisotropy antiferromagnet (AF) in the correlated spin liquid state. The active magnetic layer is incorporated into a magnetoresistive heterostructure such as MTJ [Fig.~\ref{fig:schematic}(a)], enabling its control by STT and readout via TMR. We also discuss practical device limitations and benefits.

STT-driven magnetization dynamics of a nanoscale single-domain F is described by the macrospin Landau-Lifshitz-Gilbert-Slonczewski (LLGS) equation,
 
 \begin{equation}\label{eq:LLGS}
 \begin{split}
 \frac{d\hat{m}}{dt}=-\mu_{0}\gamma\hat{m}\times\vec{H}_{eff}-\alpha\hat{m}\times\frac{d\hat{m}}{dt}\\
 +\sigma_{DL}\hat{m}\times(\hat{m}\times\hat{s})+\sigma_{FL}\hat{m}\times\hat{s}.
 \end{split}
 \end{equation}
 Here, $ \hat{m} $ and $ \hat{s} $ are unit vectors along the magnetizations of the free layer and the polarizer, respectively [Fig.~\ref{fig:schematic}(a)],  $\mu_{0} $ is the vacuum permittivity, $\alpha$ is Gilbert damping, and $\gamma$ is the gyromagnetic ratio. The effective field $\vec{H}_{eff}$ includes crystalline and/or shape anisotropies, and the external and demagnetizing fields.  The damping-like (DL) and the field-like (FL) contributions to Slonczewski's STT~\cite{slonczewski1996current} are characterized by efficiencies $\sigma_{DL}$ and $\sigma_{FL}$, respectively, which are proportional to the driving current density $J$ and its spin polarization $P$. For instance, the DL-STT efficiency is $\sigma_{DL}=-\gamma\hbar JP/4M_{s}ed$, where $\hbar$ is the Planck's constant, $M_{s}$ is the saturation magnetization, and $d$ is the thickness of F.
 
 First, we consider a simple case of negligible magnetocrystalline and shape anisotropies, and $\hat{s}$ normal to the plane. In this geometry, the effective FL-STT field is normal to the plane. Its effect is negligible at practical driving currents due to the large demagnetizing field. The solution of Eq.~(\ref{eq:LLGS}) is the out-of-plane precession (OPP) mode driven by DL-STT~\cite{lee2005analytical,ebels2008macrospin,houssameddine2007spin,lim2009measurements},

 \begin{equation}
 \label{eq:OPP}
 \begin{aligned}
 \hat{m}_{z}&=\sigma_{DL}/\alpha\mu_{0}\gamma M_S\\
 \hat{m}_{x}+i{m}_{y}&= e^{i\sigma_{DL}t/\alpha+i\phi_0}\sqrt{1-m^{2}_{z}}.
  \end{aligned}
 \end{equation}
 
Since the angular velocity $\omega=\sigma_{DL}/\alpha$ of precession is inversely proportional to damping, this dynamics can be described as viscous. The absence of a damping-dependent threshold current demonstrates that efficient STT-driven dynamics can be achieved even at large $\alpha$. 

The angle between $\hat{m}$ and $\hat{s}$ remains constant, and does not produce resistance variations due to precession. To generate magnetoelectronic signals, $\hat{s}$ can be tilted towards the x-axis. We assume that the tilt angle is sufficiently small, so that the effect of tilting on the dynamics can be neglected. The angular dependence of TMR becomes $R(\varphi)=R_{0}+\Delta Rcos(\varphi)$, as illustrated in Fig.~\ref{fig:schematic}(b). The dependence $R(\varphi)$ is almost linear in the approximately $120^\circ$ range of $\varphi$ around $90^\circ$, as shown by a hatched region in Fig.~\ref{fig:schematic}(b). In this region, precession of $\hat{m}$ produces a nearly linear variation of $R$ at a rate proportional to the current, i.e., the resistance change is proportional to the total charge $Q$ that flows through the device, as expected for an ideal memristor~\cite{chua1971memristor}. The memristive response, characterized by the slope of $R(Q)$, is tunable by varying  $\alpha$.

\begin{figure}
	\includegraphics[width=\columnwidth]{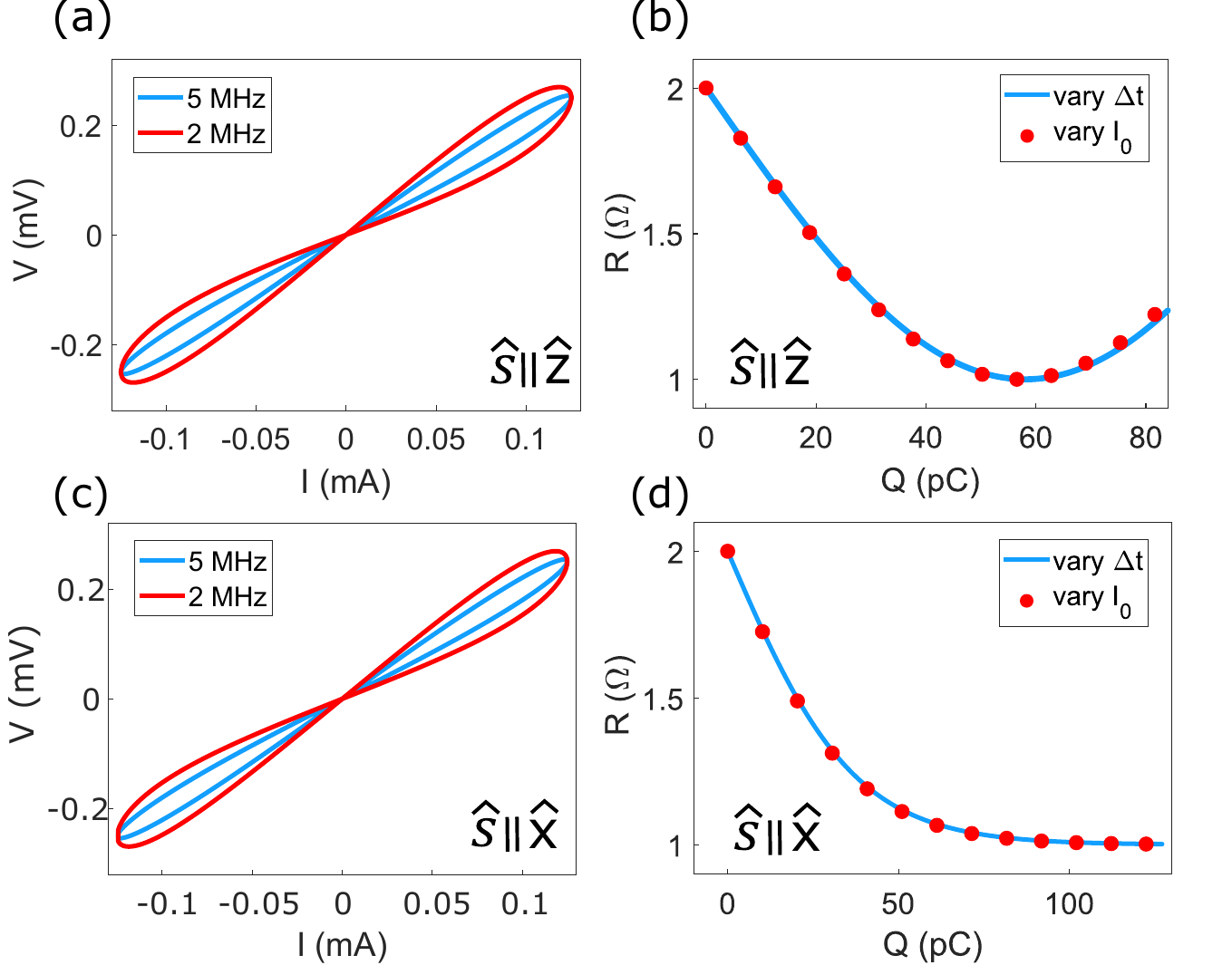}
	\caption{Results of simulations using Eq.~(\ref{eq:LLGS}) for a $5$~nm-thick, $100$~nm-diameter circular disk with $ M_{s}=800\;kA/m $, $ P=0.3$, $\alpha=10$, and negligible in-plane anisotropy. $R_{0}=2$~$\Omega$ and $\Delta R=1$~$\Omega$ were used for magnetoelectronic coefficients. (a),(c) IV hysteresis loop for $\hat{s}\parallel \hat{z}$ (a) and $\hat{s}\parallel \hat{x}$ (c), at two frequencies of sinusoidal driving current. (b),(d) Dependence of resistance variation on the total charge $Q=I_0*\Delta t$ of the current pulses with varied amplitude $I_0$ at fixed duration $\Delta t=100\;ns$ [curves], and varied $\Delta t$ at fixed $I_0=0.24\;mA$ [dots], for $\hat{s}\parallel \hat{z}$ (b) and $\hat{s}\parallel \hat{x}$ (d), with $\varphi=90^\circ$ at $t=0$.}\label{fig:ideal}
\end{figure}

Our analysis is supported by numerical integration of Eq.~(\ref{eq:LLGS}) for sinusoidal driving current, which is common in the characterization of memristors~\cite{strukov2008missing,yu2011electronic}. The pinched IV hysteresis [Fig.~\ref{fig:ideal}(a)] confirms memristive functionality. The hysteresis decreases with increasing ac current frequency $f$, as expected since the charge that flows through the device over any given part of the driving cycle is $\propto 1/f$. Figure~\ref{fig:ideal}(b) shows the calculated response to square current pulses of varied amplitude $I_0$ and duration $\Delta t$. The dependence $R(Q)$ is linear at small $Q=I_0\Delta t$, and is the same for the varied pulse amplitude as for the varied duration, confirming ideal memristive behaviors. At larger $Q$, the dependence becomes nonlinear and then non-monotonic due to $\varphi$ exceeding $180^\circ$, resulting in the loss of memristive functionality.

This problem can be overcome by utilizing in-plane polarization $\hat{s}$. To analyze this case, we consider the overdamped limit $\alpha\gg 1$. The left-hand side of Eq.~(\ref{eq:LLGS}) is negligible compared to the Gilbert damping term. Taking a cross-product of Eq.~(\ref{eq:LLGS}) with $\hat{m}$, we obtain,
\begin{equation}
\alpha\frac{d\hat{m}}{dt}=-\mu_{0}\gamma\vec{H}_{eff}+\sigma_{DL}\hat{m}\times\hat{s}+\sigma_{FL}\hat{s}.
\label{eq:overdamped}
\end{equation}
This equation describes a rotation of $\hat{m}$ towards the net effective field comprising $\vec{H}_{eff}$, the effective DL-STT field $\vec{H}_{DL}=-\sigma_{DL}\hat{m}\times\hat{s}/\mu_{0}\gamma$, and the effective FL-STT field $\vec{H}_{FL}=-\sigma_{FL}\hat{s}/\mu_{0}\gamma$. For in-plane $\hat{s}$, $\vec{H}_{DL}$ slightly tilts $\hat{m}$ out of plane, which can be neglected. Meanwhile, $\vec{H}_{FL}$ rotates $\hat{m}$ towards the orientation parallel or antiparallel to $\hat{s}$, depending on the sign of $\sigma_{FL}$. 

The magnitude of FL-STT depends on the system's geometry and its electronic properties, beyond the scope of this work. Here, we for simplicity assume $\sigma_{FL}=\sigma_{DL}$. Figure~\ref{fig:ideal}(c) shows that the IV curves exhibit the same hysteretic features as for $\hat{s}\parallel\hat{z}$. The dependence $R(Q)$ is linear at small $Q$, confirming ideal memristor functionality [Fig.~\ref{fig:ideal}(d)]. At large $Q$, the dependence $R(Q)$ saturates as $\hat{m}$ approaches $\hat{s}$, but does not become non-monotonic [Fig.~\ref{fig:ideal}(d)]. We note that saturation is expected for all memristors, otherwise their resistance would diverge or become negative. As an additional benefit of this configuration, TMR provides a more efficient readout of the memristor state. This mode of memristor operation can be also accomplished using spin-orbit torques, which can be described by in-plane $\hat{s}$~\cite{pai2014enhancement,sinova2015spin}.

\begin{figure}
	\includegraphics[width=\columnwidth]{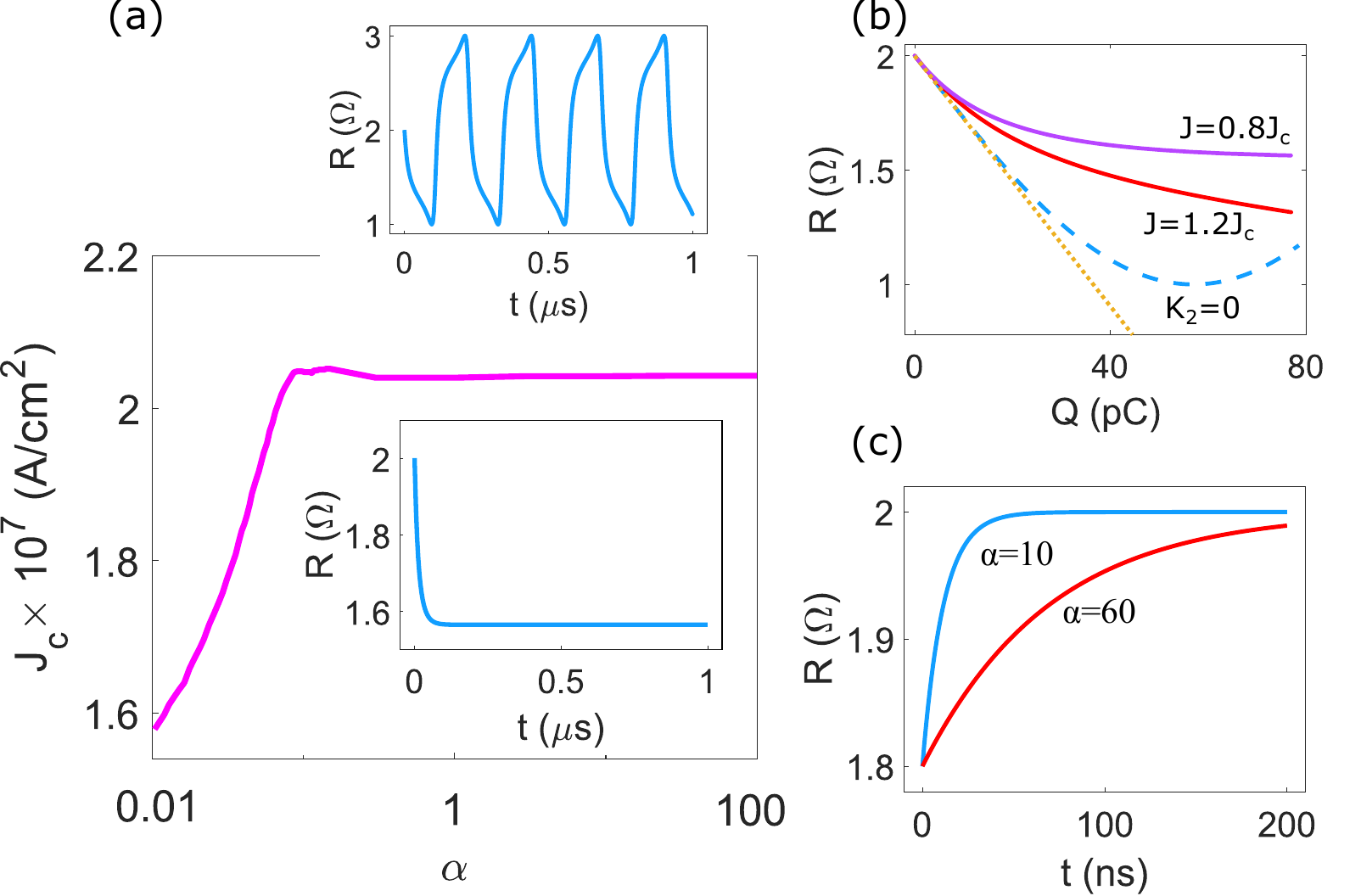}
	\caption{Effects of finite in-plane anisotropy. (a) Critical current density $J_c$ for the onset of OPP mode vs $\alpha$, for $ K_2=4$~kA/m. Insets: $R(t)$ for $J=0.8J_C$ (bottom), and $J=1.2J_C$ (top), at $\alpha=10$. (b) $R(Q)$ for square current pulses with varied duration, at $J=0.8J_C$ and $1.2J_C$. The line shows the ideal memristive behavior, the dashed curve shows the result for negligible anisotropy. (d) $R(t)$ for $\varphi_0=20^{\circ}$, at the labeled values of $\alpha$.
	}\label{fig:anisotropic}
\end{figure}

We now analyze the effects of finite in-plane anisotropy, always present in real systems due to imperfections. For $\hat{s}\parallel\hat{z}$, OPP is expected to onset above some critical current density $J_C$ at which STT overcomes the anisotropy. Consider uniaxial  anisotropy characterized by anisotropy coefficient $K_2$ and easy $y$-axis. At $J<J_C$, STT-driven rotation of $\hat{m}$ is expected to stop at  $\varphi_0=sin^{-1}(2\sigma_{DL}/\mu_0\gamma K_{2})/2$, obtained from Eq.~(\ref{eq:LLGS}) using $d\hat{m}/dt=0$.  The maximum possible value of $\varphi_0=45^\circ$, at  $\sigma_{DL}=\mu_{0}\gamma K_{2}/2$, corresponds to STT balancing the maximum anisotropy torque~\cite{ebels2008macrospin}. We note that OPP can onset at smaller $\sigma_{DL}$ than obtained from the static balance, because the trajectory of $\hat{m}$ on the Bloch sphere need not pass through the attractor at $\varphi_0=45^\circ$. Indeed, $J_C$ obtained by numeric integration of Eq.~(\ref{eq:LLGS}) slightly decreases at small $\alpha$ [Fig.~\ref{fig:anisotropic}(a)]. At large $\alpha$ relevant to this work, $J_C$ saturates to a value consistent with the above analysis, and only slightly larger than its $\alpha=0$ limit, confirming the possibility to efficiently drive viscous magnetization dynamics by STT. 

In response to square current pulses, $R(Q)$ deviates from the ideal memristive behaviors
more strongly than for the system with negligible anisotropy, due to the dependence of the anisotropy-induced torque on $\varphi$ [Fig.~\ref{fig:anisotropic}(b)]. In the subcritical regime, the magnetization can rotate only up to $\varphi_0$, limiting the operational range. However, this regime may be beneficial for avoiding the non-monotonic behaviors that compromise memristive functionality. Another consequence of anisotropy is the relaxation of $\hat{m}$ towards the easy axis, resulting in memory loss at a rate dependent on damping [Fig.~\ref{fig:anisotropic}(c)]. Using a trial solution $m_x=sin(Aexp(-t/t_1)) $, $ m_y=cos(Aexp(-t/t_1)) $, $ m_z=const$ of Eq.~(\ref{eq:LLGS}) at $I=0$, with constant relaxation time $t_1$ and $A$, we obtain $t_1=\alpha/K_2\mu_0\gamma$ for small deviations from equilibrium. 

Memory loss can be minimized by utilizing a system where damping is very large in the absence of driving, but becomes smaller at finite current. This can be accomplished by interfacing the active F layer with an ultrathin low-anisotropy antiferromagnet (AF). Exchange frustration at F/AF interface has been shown to result in the formation of a correlated spin glass at low temperatures, and viscous spin liquid above the glass transition temperature $T=T_g$~\cite{binder1986spin,ma2018spin}, imparting a large damping $\alpha\propto\nu$ to F~\cite{urazhdin2019magnetic}. The value of $\nu$ was shown to exponentially increase at $T$ approaching  $T_g$, with the latter tunable by the thickness of AF~\cite{ma2016thickness}.

Thanks to these properties, $T$ can be used as the second memristor control parameter, with $T_g$ tuned to achieve a large $\alpha$ in the absence of driving. Driving current increases the nanodevice temperature due to Joule heating, decreasing  $\alpha$ and enabling STT-driven rotation of $\hat{m}$. The variation of $\alpha$ is affected by the temporal profile of the driving current, on the timescale of the thermal relaxation time $\tau$. Below, we show that this dependence enables additional second-order memristive functionality beneficial for neuromorphic applications~\cite{kim2015experimental,wang2012synaptic}.

\begin{figure}
	\includegraphics[width=\columnwidth]{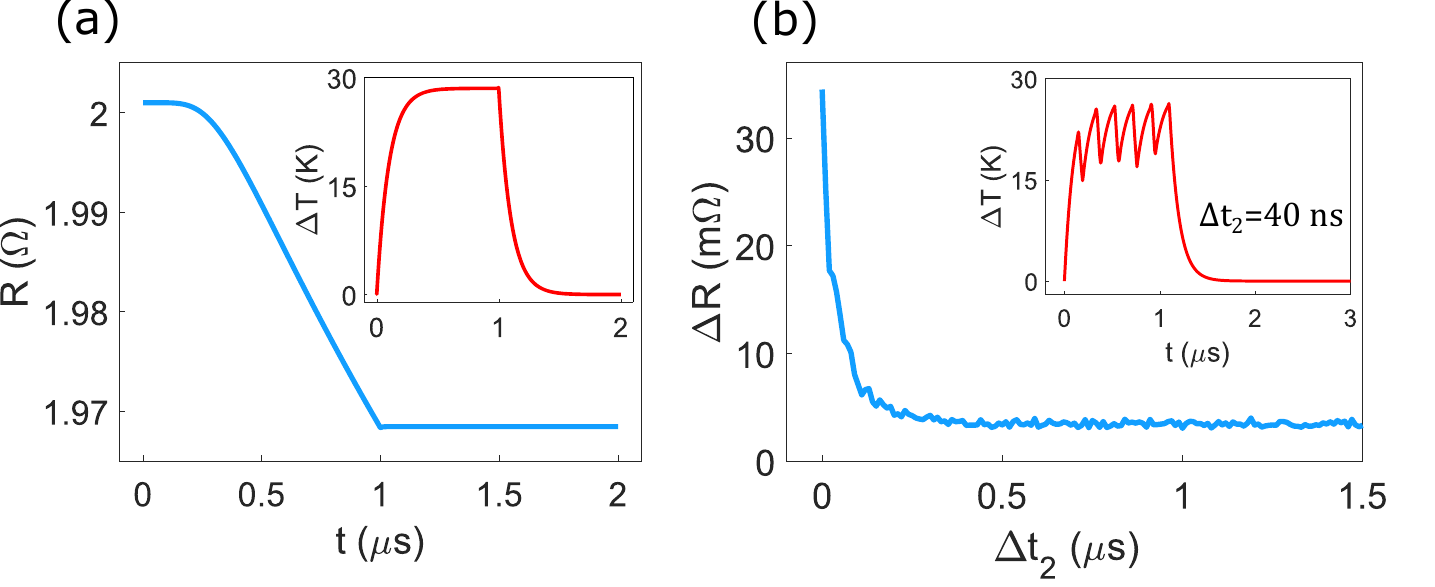}
	\caption{Effects of temperature-dependent damping in a spin glass-forming heterostructure. 
		(a) $R$ vs $t$ for a $1$~$\mu$s-long pulse of current $I_0=0.4$~mA.  Inset: $\Delta T=T-T_0$ vs $t$.  (b) Dependence of the resistance change driven by five $150$~$n$s-long pulses of current $I_0=0.4$~mA with the interval $\Delta t_2$ between the pulses. Inset: $T$ vs $t$ for $\Delta t_2=40$~ns.}\label{fig:spinglass}
\end{figure}

To model these effects, we use the relaxation time approximation for the device temperature, $dT/dt=I^2 R/C-(T-T_0)/\tau$, where $C$ is the heat capacity, $\tau$ is the thermal relaxation time, and $T_0$ is the temperature of the environment, and $\alpha(T)=\alpha(T_0)e^{a(T-T_0)}$ for the temperature dependence of damping, with $a=0.5$~\cite{urazhdin2019magnetic}.

Figure~\ref{fig:spinglass}(a) shows the time dependence of resistance in response to a square pulse of current. The pulse increases $T$ by $30$~K, as shown in the inset. The resulting decrease of damping by 3 orders of magnitude enables STT-driven rotation of $\hat{m}$ by the pulse. The temperature decreases back to $T_0$ after the end of the pulse, resulting in negligible relaxation of $\hat{m}$ towards the easy axis.

Because of the finite thermal relaxation time, the response to multiple current pulses depends on their relative timing, as illustrated in Fig.~\ref{fig:spinglass}(b) for sequential pulses. For small delay between the pulses, the temperature increase driven by a pulse remains significant at the onset of the subsequent pulse [inset in Fig.~\ref{fig:spinglass}(b)], resulting in a larger response to the latter. Thus, current-driven temperature variation serves as an additional state control variable, which provides second-order memristive functionality - dependence of the final device state on the timing of the driving current pulses. 

This feature can be particularly beneficial for synaptic functions. In real neural systems, the synaptic plasticity is controlled by the relative timing of the voltage spikes, which is realized by the sensitivity modulation mediated by the spike-driven Ca$^{2+}$ ion concentration~\cite{shouval2002unified,zucker1999calcium}. The release and the spontaneous decay of Ca$^{2+}$ results in the effective summation of closely timed pulses, leading to  spike timing-dependent plasticity (STDP), which plays an important role in the neural network's functionality. The relaxation of temperature in the proposed memristive devices closely emulates the decay of Ca$^{2+}$ concentration in neural networks, providing a simple and efficient implementation of STDP.

In summary, we have demonstrated the possibility to implement nearly ideal memristive behaviors characterized by a linear dependence of a two-terminal device resistance on the charge that flows through it, by utilizing spin torque-driven viscous magnetization dynamics. The latter property can be achieved by utilizing a spin glass-forming hybrid heterostructure with frustrated exchange interactions, which also enables non-volatile operation and the implementation of spike timing-dependent plasticity with non-overlapping pulses, due to Joule heating and the strong dependence of damping on temperature. Together with the expected high-endurance of magnetism-based memristor implementation, these characteristics make the proposed devices particularly attractive for the hardware implementation of synaptic functions in artificial neural networks.

The demonstrated direct relationship between viscous dynamics and ideal memristive behaviors can enable other efficient memristor implementations. For example, consider a dc electric motor whose shaft is mechanically attached to the rotor of a rotary rheostat. The motor and the rheostat are also electrically connected in series. For a well-lubricated motor with low inertia, the rotation angle is proportional to the charge that flows through it, i.e., its rotational dynamics is dominated by the viscous frictional losses. The resistance of the rheostat and motor connected in series is proportional to the rotation angle, and thus to the charge that flows through it, realizing a simple ideal memristor. 

This work was supported by the NSF awards ECCS-1804198 and ECCS-2005786.

\bibliography{bibl_mem}{}
\bibliographystyle{apsrev4-1}

\end{document}